\documentclass[prodmode,hillsideplop]{./utils/acmlarge}

\makeatletter
\newcommand\btxurlfont{}
\let\btxurlfont=\@empty
\newcommand\btxurldatefont{}
\let\btxurldatefont=\@empty
\makeatother
\usepackage{flafter}
\usepackage[hidelinks]{hyperref}
\usepackage[hyphenbreaks]{breakurl}
\usepackage{todonotes}
\usepackage{cleveref}
\usepackage{multicol}

\newcommand{\combref}[1]{%
  {%
    \crefformat{page}{p. ##2##1##3}%
    (\Cref{#1}, \cpageref{#1})%
    \crefformat{page}{page ##2##1##3}%
  }%
}

\acmVolume{28}
\acmYear{2021}
\acmMonth{06}

\title{Patterns for Documenting Open Source Frameworks}
\author{JOÃO SANTOS \affil{Faculty of Engineering, University of Porto. Porto, Portugal}\ \\
FILIPE F. CORREIA \affil{Faculty of Engineering, University of Porto. INESC TEC. Porto, Portugal}}

\begin{abstract}
Documenting frameworks provides its users and maintainers useful information on that software's architecture, design, and customization. Despite documentation's importance, the process of creating and maintaining it is considered to imply considerable effort, to be tedious, and expensive.

In this work, we mine patterns from open source frameworks to uncover good solutions used to document them that had not yet been described as patterns. This process resulted in four new patterns. 

\textsc{Contribution Guidelines} helps developers to become contributors to a project, helping them follow the good practices that have been adopted by its maintainers. \textsc{Documentation Versioning} consists of having separate documentation for older versions of the framework, to answer needs of the users on such versions. \textsc{Migration Handbook} helps users migrating from previous versions of the framework to newer ones. \textsc{Multi-language Support} allows translated documents in several languages to support a wider range of users for the framework.
\end{abstract}


\acmformat{Santos, J.\ and Correia,  F.F. 2021. Patterns for Documenting Open Source Frameworks.}

\copyr{PLoP'21, October 04--07, Virtual Event, {USA}. Copyright 2021 is held by the author(s). {HILLSIDE} 978-1-941652-00-8}

\begin{document}

\begin{bottomstuff}
%Permission to make digital or hard copies of all or part of this work for personal or classroom use is granted without fee provided that copies are not made or distributed for profit or commercial advantage and that copies bear this notice and the full citation on the first page. To copy otherwise, to republish, to post on servers or to redistribute to lists, requires prior specific permission. A preliminary version of this paper was presented in a writers' workshop at the 28th Conference on Pattern Languages of Programs (PLoP).
\vspace{1cm}
\end{bottomstuff}

\maketitle

\section{Introduction}

Framework documentation can be invaluable for its users, and can address different aspects, such as its system design and customization points. Despite its value, the process of creating it is often neglected, for being costly and time-consuming.

Creating and maintaining documentation for open source frameworks comes with its own peculiarities, that stem from the distributed, volunteer-based process that is typical in open source software development. The patterns that we describe in this work assume certain dynamics between the different stakeholders of open source projects. The \textit{users} of a project are those directly using and benefiting from what it offers; its \textit{contributors} are those that provide new or improved code and documentation to the project, but depend on maintainers to review and incorporate such contributions; and its \textit{maintainers} are those responsible for running the project, setting the overall vision of the project, reviewing and merging contributions, and often add and improving the project themselves. Projects often start with just one maintainer (\textit{i.e.}, the creator of the project). From an initial base of users the project main recruit new contributors, and from the most dedicated contributors recruit new maintainers.

We started this work by finding what patterns had already been written for documenting open source frameworks. We then relied on our personal experience in using and contributing to open source frameworks to analyze the documentation of five popular real-world open source frameworks, with the goal of identifying good solutions that they currently adopt, and document them as patterns. 
We searched for any recurring types of documents being used for this kind of documentation and considered specifically some issues found in the documentation process. In this process, we took inspiration from works from Aghajani et al. who, through empirical studies, identify a set of types of documentation perceived as useful by practitioners~\cite{practitioners_perspective2020} and a set of common documentation issues~\cite{documentation_issues_2019}.

This was the starting point for the pattern mining process, which resulted in the four patterns for documenting frameworks that we describe in this article.

With this work, we aim to help mainly framework contributors and maintainers to design the documentation for their projects, and indirectly also the users for those frameworks, who will consequently benefit from having effective documentation. 

\section{Related Work} \label{related_work}

Some authors propose solutions for writing and maintaining framework documentation~\cite{ademar_framework_patterns,johnson1992documenting,osterbye1999minimalist,butler1997documenting}. However, as frameworks evolve, new concerns can emerge in the process of writing and maintaining its documentation.

\begin{figure}[htbp]
  \begin{center}
    \leavevmode
    \includegraphics[width=0.5\textwidth]{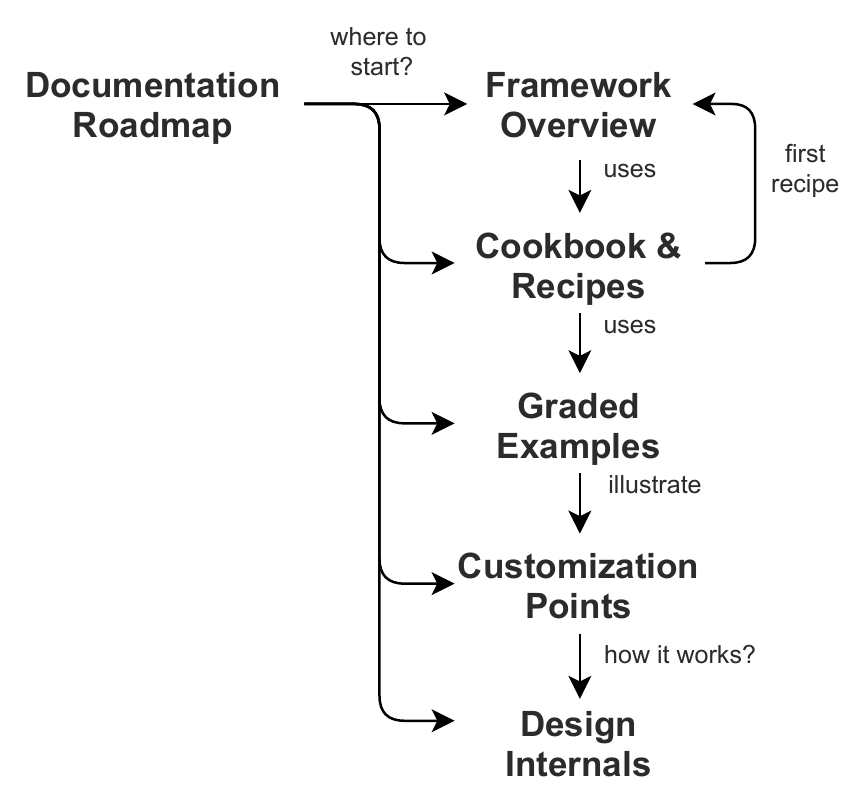}
    \caption{Patterns for documenting frameworks and their relationships}
    \label{fig:ademar_pattern_map}
  \end{center}
\end{figure}

In previous work, we surveyed literature for software documentation patterns and found 114 different documentation patterns~\cite{santos2020review}. The book \textit{Living Documentation}~\cite{martraire2019living} covers also quite a few patterns for improving software documentation with minimal cost, by relying on automation and the notion of constantly updated documentation artifacts. However, we have found no patterns with the specific purpose of describing how to document \textit{open source} frameworks. An important work for us is Aguiar's, who proposes patterns for addressing \textit{framework} documentation~\cite{ademar_framework_patterns}. We choose to look into these patterns in more detail here, as they are the closest ones to the topic we address. The six patterns that this work describes have the goal of guiding the choice of the kinds of documents to produce, how to connect them, and of which contents to include. They are depicted in Figure~\ref{fig:ademar_pattern_map} and described below.

\begin{itemize}
    \item[$\bullet$] \textsc{Documentation Roadmap} directs readers of different audiences on what they may want to read to find the knowledge they need. It is usually the main entry point to the documentation;

    \item[$\bullet$] \textsc{Framework Overview} introduces the framework. It briefly describes the general context of the framework and several of its aspects, such as the domain, goals, and kind of flexibility that it offers. 

    \item[$\bullet$] \textsc{Cookbook \& Recipes} explain how the framework can be used to address different concerns that may appear when developing a software system. It is organized as a list of recipes organized as a guide (cookbook);
    
    \item[$\bullet$] \textsc{Graded Examples} describe examples of how the framework can be used, from simple to complex ones, linking them with other forms of documentation.
    
    \item[$\bullet$] \textsc{Customization Points} describes the framework's customization points (hot-spots) and how they may be put into practice. They often include examples;
    
    \item[$\bullet$] \textsc{Design Internals} provide information of the system design, and especially of its hot-spots, explaining the general rationale for the design, and what and how can be adapted to different purposes;
\end{itemize}

\section{Patterns Overview} \label{patterns_overview}

\begin{figure}[htbp]
  \begin{center}
    \leavevmode
    \includegraphics[width=1\textwidth]{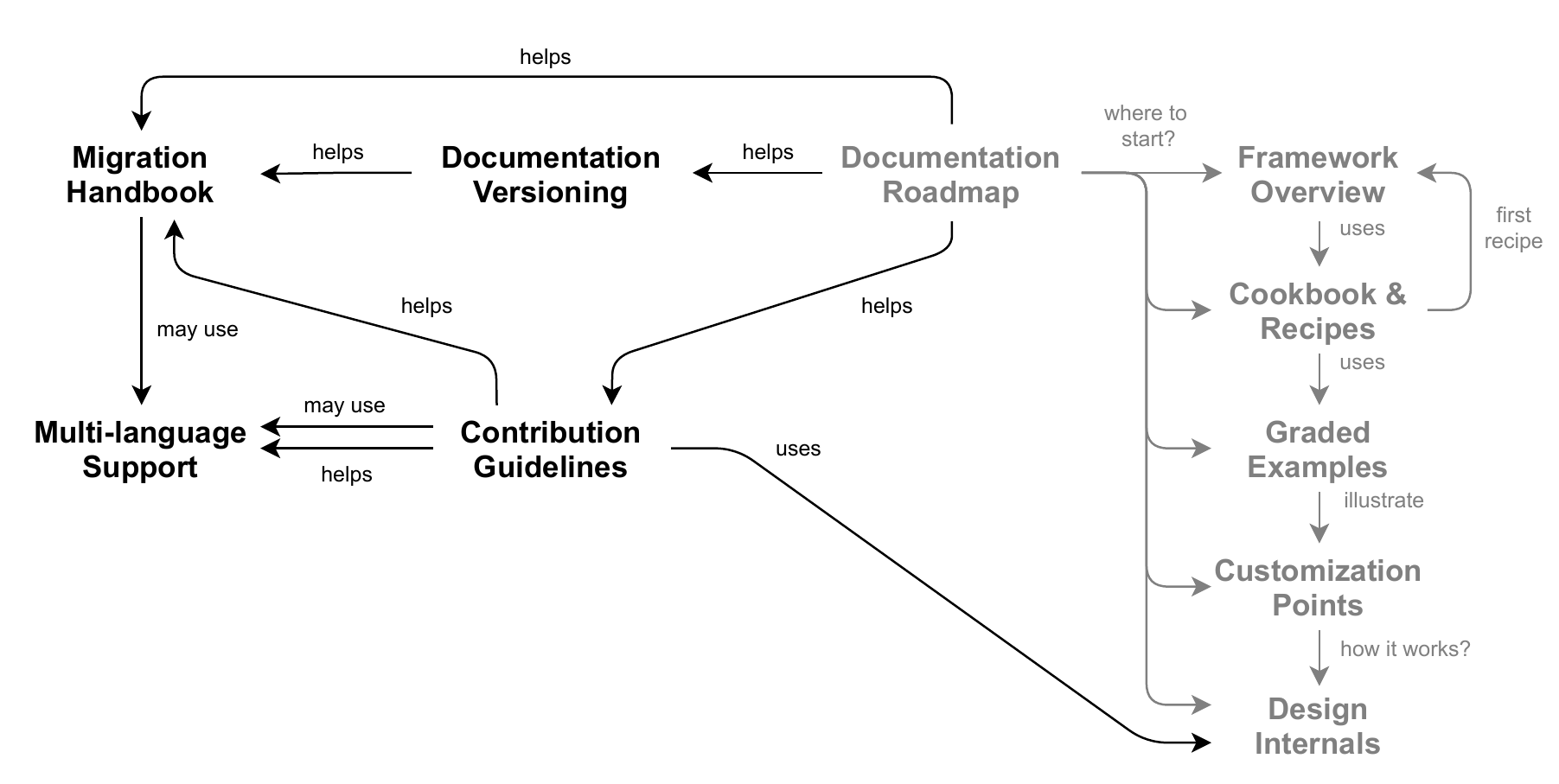}
    \caption{Patterns for documenting open source frameworks and their relationships}
    \label{fig:pattern_mining_relationship}
  \end{center}
\end{figure}

We identify and describe four patterns: \textsc{Contribution Guidelines}, \textsc{Documentation Versioning}, \textsc{Migration Handbook}, and \textsc{Multi-language Support}. They are, with their relationships, depicted in Figure~\ref{fig:pattern_mining_relationship}. While \textsc{Contribution Guidelines} and \textsc{Migration Handbook} are about two specific types of documents that appear while documenting frameworks, \textsc{Documentation Versioning} and \textsc{Multi-language Support} can be said to be more about how to maintain and navigate specific kinds of documentation.

Receiving contributions is an essential part of open source projects, and project maintainers often encourage \textit{users} to become \textit{contributors}. \textsc{Contribution Guidelines} facilitates new contributions by giving relevant information on how contribution runs for that particular project. This is especially relevant for frameworks, as the users of software frameworks are usually \textit{software developers} and, therefore, are in a better position to become contributors than users from other types of open source projects.

Despite the release of new framework versions, older versions may still be in use for some time, and the documentation for them should be kept available. \textsc{Documentation Versioning} supports that by providing access to the documentation for the different versions.

New software versions can introduce changes on how to implement framework features, which can make it challenging to upgrade from a previous framework version. \textsc{Migration Handbook} handles those challenges by informing framework users on how to operate these changes.

The language in which the documentation is written should not be a barrier for framework users. \textsc{Multi-language Support} eases that barrier by offering documentation content in different languages.

\section{Pattern Form} 
\label{sec:pattern_form}

Several authors define their patterns considering different forms~\cite{pattern_language_alexander,gamma1995elements,posa_vol_1,ademar_framework_patterns,filipe_consistent_patterns,filipe_correia_phd}, which we have considered to find an adequate one to present our patterns. For this paper we adopt the following structure:

\begin{itemize}
    \item \textbf{Name.} The pattern name conveys the key aspects of its solution is about.
    
    \item \textbf{Introductory paragraph.} This part of the pattern describes the context in which the problem emerges and the solution is applied.
    
    \item \textbf{Problem.} This section begins by defining an overview of the problem and is followed by a more detailed description where the forces are identified.
    
    \item \textbf{Solution.} This section starts by presenting an initial statement of the pattern solution. Then it goes through the solution with more details, mentioning the consequences of the pattern when relevant.
    
    \item \textbf{Related Patterns.} This section explains which patterns relate with the one being described and how that relationship proceeds.
    
    \item \textbf{Known Uses.} This section describes real-world examples of where the pattern is adopted.
\end{itemize}

\section{\textsc{Contribution Guidelines}} \label{contribution_guidelines} 

You are designing the documentation of an open source framework. One aspect in which open source projects often differ from commercial ones, is that they live from volunteer work from people spread around the world, that self-organize towards a common goal. As a maintainer of a project, you may feel the need to gather more contributors from the day that the first version of a project is made available. 

\subsection{Problem}
\textbf{How to help users contribute to a project?}

As the project evolves in size and complexity, more users are likely to want to become contributors. However, to do that they may need to know more about the system internals, how to report bugs, and which code conventions are adopted by the team.

As more users want to contribute to an open source project, the ability of project maintainers to on-board them is strained. Projects taking their first steps often provide such support through mailing lists or text chats, which allow maintainers to address the questions that potential contributors may have, in all their specificity, but this is a solution that is hard to scale to a large number of contributors, given the effort that it implies.

It also allows inconsistencies, as different maintainers may provide different answers to contributors' questions as they surface. This calls for a single source of truth of how to contribute to a project, but creating such a permanently updated and comprehensive content may also imply a considerable effort. 

\subsection{Solution}
\textbf{Provide guidelines on how to contribute to the project and which best practices the team follows.}

Start by creating a document, or set of documents, that provide enough information on how to contribute to the project. Information commonly found in \textsc{Contribution Guidelines} includes code conventions that the team is following, how to report bugs, and how to submit changes, but each project should define what directions should its \textsc{Contribution Guidelines} include, based on the most frequent doubts and mistakes done by new contributors. The \textsc{Contribution Guidelines} can also point to an overview of the system to let the user know, early, what are the main building blocks of the system.

Make sure that potential contributors can easily find these documents. One way to do that is to point to it from a \texttt{README} file, which often includes the \textsc{Documentation Roadmap}~\cite{ademar_framework_patterns}. Also, avoid having some of this information appear in other parts of the documentation, trying to ensure that all the information on how to contribute to the project is gathered in this document. 

Creating and maintaining this document is bound to require more effort than answering a specific question from a new contributor, or than reviewing a particular contribution to the project, but that effort will easily payoff as many potential contributors can benefit from having this document available.

\subsection{Related Patterns}
\begin{itemize}
    \item[$\bullet$] Guidelines should describe the design principles of the system. Using \textsc{The Big Picture}~\cite{book_patterns_agile_documentation}~\combref{app:the_big_picture} allows readers to understand the overall system architecture.
    
    \item[$\bullet$] Authors want to help contributors on how to understand better their systems. Considering that, \textsc{Design Internals}~\cite{ademar_framework_patterns}~\combref{app:design_internals} is used to describe the fundamental layers behind the framework, and potentially the areas where customization is possible.
    
    \item[$\bullet$] Often, contributors don't know how they should read the guidelines, considering that we may use the \textsc{Guidelines for Readers}~\cite{book_patterns_agile_documentation}~\combref{app:guidelines_for_readers} to introduce an initial overview towards documents that help contributors facing these documents.
    
    \item[$\bullet$] Knowing how to reach this section is also important for users. \textsc{Documentation Roadmap}~\cite{ademar_framework_patterns}~\combref{app:documentation_roadmap} helps by organizing all documentation sections, and in particular including this contributing section in their organization.
    
    \item[$\bullet$] \textsc{Contribution Guidelines} may use \textsc{Multi-language Support} to provide translated guidelines for users who don't speak the  language in which guidelines may have been originally written.
\end{itemize}

\subsection{Known Uses}
\begin{itemize}
    \item[$\bullet$] \textbf{React.js}~\footnote{\url{https://reactjs.org/docs/how-to-contribute.html}} implements \textsc{Contribution Guidelines} by containing one main page that includes several explanations on how to contribute to the code and documentation, their design principles, code of conduct, versioning policy, and style guide.
    
    \item[$\bullet$] \textbf{Gatsby}~\footnote{\url{https://www.gatsbyjs.com/contributing/}} uses this pattern by providing one main page with six different sections that explain their contribution process to code, blog, and documentation, followed by their brand guidelines, and it includes a description of the community behind the project.
    
    \item[$\bullet$] \textbf{Angular}~\footnote{\url{https://angular.io/contribute}} adopts \textsc{Contribution Guidelines} by providing a page with all Angular projects. The main platform link redirects to a markdown file in GitHub. The file contains information about the code of conduct, how to report bugs, their submission guidelines, the review process, and the code style.
\end{itemize}

\newpage
\section{Documentation Versioning} \label{documentation_versioning} 
You are designing the documentation of an open source framework. With the launch of each new version of the framework, documentation is usually updated to reflect the API and features of the new version. However, not all frameworks users will adopt the last version, some will keep using previous versions.

\subsection{Problem}
\textbf{How can users access previous versions of the documentation?}

While new versions of your framework are launched, some users will keep using older versions and need to access their documentation, navigating and searching it as needed, as they do for the latest version. It's also needed that the documentation is kept up-to-date with the latest version of the framework, but also that it is possible to update the documentation for older versions of the framework to incorporate fixes to the documentation itself, or to reflect new minor versions that need to be released for older framework versions (\textit{e.g.}, to correct bugs or security issues).

\subsection{Solution}
\textbf{Maintain different versions of the documentation, one for each release of the framework.}

Throughout the development of a new version of the framework, work also on the new documentation for that version. When the new version of the framework is released, make its documentation available to users, and allow them to easily navigate between the different versions. To support navigating across documentation versions, provide a list of previous versions of your framework and links to the documentation for those versions. Another way to support navigation is to place a dropdown in each documentation page that allows to switch to the same page for a different version of the framework.

To reduce the effort of maintaining multiple versions of the documentation and ensure its kept updated, use tools that make it easier to compare different documentation versions and to associate them to the framework versions that they refer to. A possible approach is to keep the documentation in the same repository used to version the framework itself. Established mechanisms for versioning source-code, such as \textit{branches} and \textit{tags} can also be used for documentation and keeping both code and documentation in the same repository ensures that documentation is always close to its framework version. 

% \todo[inline]{
% \par Sugestão c/ consequências

% ...

% Even knowing that including the documentation for all versions requires more effort and time, this is easily justified by all the users who still use previous versions of your framework and benefit from having documentation accessible for that specific version.
% }

\subsection{Related Patterns}
\begin{itemize}
    \item[$\bullet$] The organization of different versions requires a history of these previous versions, for that this pattern may be applied after \textsc{Document History}~\cite{book_patterns_agile_documentation}~\combref{app:document_history} is applied.
    
    \item[$\bullet$] \textsc{Documentation Roadmap}~\cite{ademar_framework_patterns}~\combref{app:documentation_roadmap} helps \textsc{Documentation Versioning} by including an option to change between documentation versions, or to indicate the documentation version which the user is reading.
    
    \item[$\bullet$] This pattern also supports \textsc{Document Archive}~\cite{book_patterns_agile_documentation}~\combref{app:document_archive} by providing an organization that allows navigation through previous versions.
\end{itemize}

\subsection{Known Uses}
\begin{itemize}
    \item[$\bullet$] \textbf{React.js}~\footnote{\url{https://reactjs.org/versions}} adopts \textsc{Documentation Versioning} by providing a list of previous versions as well as their changelogs and snapshots.
    
    \item[$\bullet$] \textbf{Vue.js} implements this pattern by presenting a dropdown menu with previous versions as values. Each change of the dropdown redirects to the documentation version page.
    
    \item[$\bullet$] \textbf{Angular} also adopts this pattern by showing a dropdown menu with their previous versions. A redirect occurs after changing the dropdown value.
\end{itemize}

\section{\textsc{Migration Handbook}} \label{migration_handbook} 

You are designing the documentation of an open source framework. You have to take into account that frameworks evolve over time, and more and different features will likely become available with each new version. It's possible that breaking changes are introduced with newer versions of the framework, as earlier versions might not anticipate all future needs. 
%“once something is released into the wild, bugs or imperfections quickly become essential features and are nearly impossible to change” -- Brendan Eich

\subsection{Problem}
\textbf{How can users migrate systems from an older to a newer version of the framework?}

Users who use older versions of the framework may want to update due to security reasons or because the new version includes more features. However, migrating from one older version to a more recent one often implies a significant effort. For example, it is possible for users to understand how the migration is done by reviewing all the changes that the framework went through between the older and newer version, to understand how their use of the framework needs to change. Some additional guidance is useful to make the process of migrating easier and less error-prone.

\subsection{Solution}
\textbf{Provide a guide containing any change that the new version introduces and is relevant for anyone trying to upgrade to that version from the previous one.}

Start by collecting all changes to include in the \textsc{Migration Handbook}; this can be done as development progresses, or by going through the history of the source code repository before a new version of the framework is released. The \textsc{Migration Handbook} often links to the \textit{release notes} for that framework version, where the new features that it introduces are briefly described, and it introduces what users of the framework need to change in their projects to use the new version of the framework. This includes highlighting the API changes, but also any API feature that might have been made obsolete or deprecated in the new version. 
The description of each change can be extended with the motivation behind the decision to introduce it and sometimes a link to a \textit{pull request}\footnote{\textit{Pull requests} are operations supported by some project-tracking platforms,  where a contributor asks the maintainers of a repository to review a set of changes to the code and documentation, and merge the int the project.} where it was added.

Features described as \textit{obsolete} or \textit{deprecated} will usually be accompanied by a description of alternative ways to use the framework. This includes using code snippets of how it was done in the previous version and how it's done in the new version, so users can understand better those changes with examples.

% \todo[inline]{
% \par Sugestão c/ consequências

% Start by collecting all changes to include in the document. The \textsc{Migration Handbook} often links to the \textit{release notes} for that framework version, where the new features that it introduces are briefly described, and it introduces what users of the framework need to change in their projects to use the new version of the framework. This includes highlighting the API changes, but also any API feature that might have been made obsolete or deprecated in the new version. 

% ...

% However, writing this document can lead to incomplete information because a new version of the framework can have several changes to track. To minimize this situation, the tracking can be done as development progresses, or by going through the history of the source code repository before a new version of the framework is released.  
% }

\subsection{Related Patterns}
\begin{itemize}
    \item[$\bullet$] Users who use outdated versions can be notified of a significant change on the document. Considering that, \textsc{Notification upon Update}~\cite{book_patterns_agile_documentation}~\combref{app:notification_upon_update} can help this pattern by providing navigation to the migration documents that allows a version upgrade.
    
    \item[$\bullet$] \textsc{Documentation Versioning} helps \textsc{Migration Handbook} by referring changes and deprecations to that specific previous version. For instance, while a user finds a particular change in a recent version it may be helpful to navigate to that version to see more API details.
    
    \item[$\bullet$] This pattern may use \textsc{Multi-language Support} to facilitate migration for users who don't speak the original language in which the documentation was written.
    
    \item[$\bullet$] \textsc{Contribution Guidelines} helps \textsc{Migration Handbook} by defining best practices of how to describe the process of upgrading to a newer version.
\end{itemize}

\subsection{Known Uses}
\begin{itemize}
    \item[$\bullet$] \textbf{Vue.js}~\footnote{\url{https://vuejs.org/v2/guide/migration.html}} applies this pattern containing two main parts: the first is a FAQ section approaching relevant questions to their users; the second describes API changes with code snippets to help users to fulfill their goals.
    
    \item[$\bullet$] \textbf{Angular}~\footnote{\url{https://angular.io/guide/updating-to-version-12}} adopts \textsc{Migration Handbook} describing an initial overview regarding the migration, followed by the changes and deprecations of the newer version. Each change is detailed, providing a link to the PR where that change occurred.
    
    \item[$\bullet$] \textbf{ASP.NET Core}~\footnote{\url{https://docs.microsoft.com/en-us/aspnet/core/migration/31-to-50?view=aspnetcore-5.0&tabs=visual-studio}} implements this pattern by containing different versions that can be migrated, followed by the migration prerequisites. Additionally, they provide a detailed explanation for each feature, containing the old behavior, the new behavior, and in some cases, the motivation behind that decision.
\end{itemize}

\section{\textsc{Multi-Language Support}} \label{multilanguage_support} 
You are designing the documentation of an open source framework. The need to provide good documentation to all the potential users of a framework exists from the day that the first version of a project is made available. This may be easier in the early days of a project, but a framework that grows in use is more likely to have a significant number of users that doesn't speak the main language in which the documentation is written.

\subsection{Problem}
\textbf{How to remove language barriers from users who don't speak the language in which the documentation is written?}

The framework users should have as little barriers as possible for accessing and understanding its documentation. One of such barriers can be language. Even though many projects adopt English as their main language, it is not a guarantee that it makes them able to reach all of its potential users. Producing translations of the documentation to many languages could be an option, but it would imply a considerable effort to create and maintain then. This becomes even more challenging if there aren't yet in the project contributors that can write in those languages. However, contributing to translations is one of the easiest ways to start immediately contributing to a open source project.

\subsection{Solution}
\textbf{Produce and distribute documentation artifacts in the languages that are spoken by your users.}

Start by providing the mechanisms for receiving, reviewing and accepting contributions of documentation in new languages. A possible approach is to keep the documentation in the same repository used to version the framework itself, and allow the same mechanisms used to receive contributions (\textit{e.g.}, \textit{pull requests}). Encourage contributors to use as templates the documentation already produced for other languages. 

Allow to switch between different languages, possibly using a similar approach to the one used to switch between different \textsc{Documentation Versions}. If particular parts of the documentation are not available for the given language, make sure to state that explicitly to users, and allow them to switch to the documentation in one of the languages that are available.

% \todo[inline]{
% Sugestão c/ consequências

% Start by providing the mechanisms for receiving, reviewing and accepting contributions of documentation in new languages. A possible approach is to keep the documentation in the same repository used to version the framework itself, and allow the same mechanisms used to receive contributions (\textit{e.g.}, \textit{pull requests}).

% ...

% As the project evolves and multiple users add translations, this situation can lead to inconsistency between documents. One way to mitigate this is to encourage contributors to use as templates the documentation already produced for other languages.
% }

\subsection{Related Patterns}
\begin{itemize}
    \item[$\bullet$] Translated documents can be hard to found by users. For that, this pattern can be supported by \textsc{Information Marketplace}~\cite{book_patterns_agile_documentation}~\combref{app:information_marketplace} letting the users know that exists a translation in their native language for that document.
    
    \item[$\bullet$] Writing the same document with different languages can be a complex task. To ease that, \textsc{Documentation Templates}~\cite{internal_docs}~\combref{app:documentation_templates} can help this pattern by providing templates which only vary in language.
    
    \item[$\bullet$] \textsc{Contribution Guidelines} helps \textsc{Multi-language Support} by emphasizing best practices for contributors to follow when writing documentation, particularly on how they should translate documents.
\end{itemize}

\subsection{Known Uses}
\begin{itemize}
    \item[$\bullet$] \textbf{ASP.NET Core}~\footnote{\url{https://docs.microsoft.com/en-us/locale/?target=https://docs.microsoft.com/en-us/aspnet/core/?view=aspnetcore-5.0}} uses this pattern, offering a possibility in the footer to change the language in which the documentation is translated. This language change is done on an external page that redirects to the translated documentation page.
    
    \item[$\bullet$] \textbf{Angular} implements \textsc{Multi-language Support} by offering all languages available in the footer. The selection of the new language will redirect the user to the translated page.
    
    \item[$\bullet$] \textbf{React.js}~\footnote{\url{https://reactjs.org/languages}} adopts this pattern by providing a link in their menu to the translations page. That page contains all available translations that are located individually on a new page.
\end{itemize}

\section{Conclusion and Future Work} \label{conclusions}

We expect that the four patterns for documentation open source frameworks that we propose in this paper---\textsc{Contribution Guidelines}, \textsc{Documentation Versioning}, \textsc{Migration Handbook}, and \textsc{Multi-language Support}---can help maintainers and contributors of open source frameworks to adopt the documentation practices that they describe.

By formalising these practices as patterns we also expect to support the next steps in our research, which we can outline in the following way:
\begin{enumerate}
    \item It could be useful to understand if these patterns are applicable outside the strict domain of open source frameworks. Even though the patterns were mined in this specific context, we believe they can be used when documenting other kinds of software, and would find it interesting to study their presence in such other domains;
    \item Studying a larger number of open source projects could help to identify more patterns and bring further insights about the patterns that we have already documented;
    \item Lastly, and following up on initial work that we have already developed in this direction~\cite{joaosantos_master_thesis}, we propose performing an empirical study of the adoption of these and other framework patterns, seeking insights on the contexts under which the patterns are most useful. We would use the results to produce a guide that framework maintainers and contributors could use when writing and maintaining documentation.
\end{enumerate}

\section{Acknowledgments} \label{acknowledgments}

We would like to show our appreciation for Eduardo Guerra for his valuable feedback to preliminary versions of this work, to Dionysios Athanasopoulos for shepherding this paper for PLoP 2021, and to all the participants of the Douro writers' workshop at PLoP 2021---Apitchaka Singjai, Eduardo B. Fernandez, Jason Jaskolka, Joe Yoder, Jomphon Runpakprakun and Rebecca Wirfs-Brock. All of you, thank you for so generously contributing to the paper through discussion, and by providing insightful feedback that helped us to greatly improve our work.

\newpage
% Bibliography
\bibliographystyle{./utils/ACM-Reference-Format-Journals}
\bibliography{./others/references}

% Appendix
% Appendix
\newpage
\appendix
\section{Patlets for the related patterns} \label{app:related_patterns}

\par The related patterns mentioned in the literature are described below as patlets~\cite{posa_vol_1}. Each patlet contains the pattern name, the problem, and the solution.

\begin{multicols}{2}

\begin{center}
\par \textbf{\textsc{Cookbook and Recipes}}~\cite{ademar_framework_patterns} \label{app:cookbook_and_recipes}
\end{center} 
\par \textbf{Problem:} ``How to quickly provide users with information that helps them learning how to use the framework?''
\par \textbf{Solution:} ``Provide a collection of recipes, one for each framework customization, organized in a cookbook, which acts as a guide to the contents of all its recipes.''

\begin{center}
\par \textbf{\textsc{Customization Points}}~\cite{ademar_framework_patterns} \label{app:customization_points}
\end{center} 
\par \textbf{Problem:} ``How to help readers learn in detail how to customize a specific part of a framework?''
\par \textbf{Solution:} ``Provide a list of the framework’s customization points, also known as hot-spots, i.e., the points of predefined refinement where framework customization is supported, and, for each one, describe in detail the hooks it provides and the hot-spot subsystem that implements its flexibility.''

\begin{center}
\par \textbf{\textsc{Design Internals}}~\cite{ademar_framework_patterns} \label{app:design_internals}
\end{center} 
\par \textbf{Problem:} ``How to help framework users on quickly grasping the design and implementation of a framework to support them on achieving advanced customizations, not typical, or not specifically documented?''
\par \textbf{Solution:} ``Provide concise detailed information about the design internals of the framework, especially the areas designed to support configuration, known as hot-spots. You should start by describing how the framework hot-spots support configuration.''

\begin{center}
\par \textbf{\textsc{Document Archive}}~\cite{book_patterns_agile_documentation} \label{app:document_archive}
\end{center} 
\par \textbf{Problem:} ``How can projects avoid the loss of any document versions?''
\par \textbf{Solution:} ``Archiving project documentation offers the advantage that versions can be retrieved when necessary.''

\begin{center}
\par \textbf{\textsc{Document History}}~\cite{book_patterns_agile_documentation} \label{app:document_history}
\end{center} 
\par \textbf{Problem:} ``How can confusion be avoided between versions of a document?''
\par \textbf{Solution:} ``A document history can explain the differences to previous versions of a document, and can relate the document to versions of the software it describes.''

\begin{center}
\par \textbf{\textsc{Documentation Roadmap}}~\cite{ademar_framework_patterns} \label{app:documentation_roadmap}
\end{center} 
\par \textbf{Problem:} ``How to help readers on quickly finding in the overall documentation their way to the information they need?''
\par \textbf{Solution:} ``Start by providing a roadmap for the overall documentation, one that reveals its organization, how the pieces of information fit together, and that elucidates readers of different audiences about the main entry points and the paths in the documentation that may drive them quickly to the information they are looking for, especially at their first contact.''

\begin{center}
\par \textbf{\textsc{Documentation Templates}}~\cite{internal_docs} \label{app:documentation_templates}
\end{center} 
\par \textbf{Problem:} ``Different  programmers  have  different  styles  when  writing  and  structuring documentation.   Depending   on   which   programmer   wrote   what   in   the documentation,  some  aspects  of  the  program  are  well  explained  whereas others  are  not  -  how  can  we  ensure  some  kind  of  uniformity  of  the documentation?''
\par \textbf{Solution:} ``Identify   overall   aspects   that   should   (or   can)   be   addressed   in   the documentation and create templates for these.''

\begin{center}
\par \textbf{\textsc{Framework Overview}}~\cite{ademar_framework_patterns} \label{app:framework_overview}
\end{center} 
\par \textbf{Problem:} ``How to help readers on getting a quick, but precise, first impression of a new framework?''
\par \textbf{Solution:} ``Provide an introductory document, in the form of a framework overview, that describes the domain covered by the framework in a clear way, i.e. the application domain and the range of solutions for which the framework was designed and is applicable.''

\begin{center}
\par \textbf{\textsc{Graded Examples}}~\cite{ademar_framework_patterns} \label{app:graded_examples}
\end{center} 
\par \textbf{Problem:} ``How to help readers on getting started fast using a framework both for simple and complex kinds of usage?''
\par \textbf{Solution:} ``Provide a small but representative graded set of training examples to illustrate the framework’s applicability and features, each one illustrating a single new way of customization, smoothly growing in complexity, and eventually altogether providing complete coverage.''

\columnbreak
\begin{center}
\par \textbf{\textsc{Guidelines for Readers}}~\cite{book_patterns_agile_documentation} \label{app:guidelines_for_readers}
\end{center} 
\par \textbf{Problem:} ``How can potential readers be informed whether they should read a document, and if so, on which parts they should focus?''
\par \textbf{Solution:} ``Some brief guidelines at the beginning of each document can inform potential readers of the purpose the document serves and explain how different groups of readers should approach the document.''

\begin{center}
\par \textbf{\textsc{Information Marketplace}}~\cite{book_patterns_agile_documentation} \label{app:information_marketplace}
\end{center} 
\par \textbf{Problem:} ``How can good documents be prevented from going sadly unnoticed?''
\par \textbf{Solution:} ``Documents gain more attention if the intended readers are actively invited to read them.''

\columnbreak
\begin{center}
\par \textbf{\textsc{Notification upon Update}}~\cite{book_patterns_agile_documentation} \label{app:notification_upon_update}
\end{center} 
\par \textbf{Problem:} ``How can readers be prevented from using outdated versions?''
\par \textbf{Solution:} ``Whenever there is a significant change in a project document, all potential readers should be notified of the new version. The notification should roughly explain what has been changed, but should not include the updated material itself.''

\begin{center}
\par \textbf{\textsc{The Big Picture}}~\cite{book_patterns_agile_documentation} \label{app:the_big_picture}
\end{center} 
\par \textbf{Problem:} ``How can people be introduced to a project without being confronted with a deluge of technical details?''
\par \textbf{Solution:} ``A good feel for a project is best conveyed through a description of the ‘big picture’ of the architecture that underlies the system under construction.''

\end{multicols}

% History dates
% {Initial submission}{Conference Version}{Submission after having considered Writers' Workshop feedback}

\received{June 2021}{August 2021}{December 2021}

\end{document}